\documentclass[8.5pt,twoside,twocolumn]{article}
\usepackage[super,sort&compress,comma]{natbib} 
\usepackage{mhchem}
\usepackage{times,mathptm}
\usepackage{sectsty}
\usepackage{balance} 
\usepackage{graphicx} 
\usepackage{lastpage}
\usepackage[format=plain,justification=raggedright,singlelinecheck=false,font=small,labelfont=bf,labelsep=space]{caption} 
\usepackage{fancyhdr}
\begin{document}

\thispagestyle{plain}
\fancypagestyle{plain}{
\renewcommand{\headrulewidth}{1pt}}
\renewcommand{\thefootnote}{\fnsymbol{footnote}}
\renewcommand\footnoterule{\vspace*{1pt} 
\hrule width 3.4in height 0.4pt \vspace*{5pt}} 
\setcounter{secnumdepth}{5}

\makeatletter 
\def\subsubsection{\@startsection{subsubsection}{3}{10pt}{-1.25ex plus -1ex minus -.1ex}{0ex plus 0ex}{\normalsize\bf}} 
\def\paragraph{\@startsection{paragraph}{4}{10pt}{-1.25ex plus -1ex minus -.1ex}{0ex plus 0ex}{\normalsize\textit}} 
\renewcommand\@biblabel[1]{#1}            
\renewcommand\@makefntext[1]% 
{\noindent\makebox[0pt][r]{\@thefnmark\,}#1}
\makeatother 
\renewcommand{\figurename}{\small{Fig.}~}
\sectionfont{\large}
\subsectionfont{\normalsize} 

\fancyfoot{}
\fancyfoot[RO]{\footnotesize{\sffamily{1--\pageref{LastPage} ~\textbar  \hspace{2pt}\thepage}}}
\fancyfoot[LE]{\footnotesize{\sffamily{\thepage~\textbar\hspace{3.45cm} 1--\pageref{LastPage}}}}
\fancyhead{}
\renewcommand{\headrulewidth}{1pt} 
\renewcommand{\footrulewidth}{1pt}
\setlength{\arrayrulewidth}{1pt}
\setlength{\columnsep}{6.5mm}
\setlength\bibsep{1pt}

\twocolumn[
\begin{@twocolumnfalse}
\noindent\LARGE{\textbf{Limiting the valence: advancements and new perspectives on patchy colloids, soft functionalized nanoparticles and biomolecules}}
\vspace{0.6cm}

\noindent\large{\textbf{Emanuela Bianchi,$^{\ast}$\textit{$^{a,b}$} Barbara Capone,\textit{$^{a,c}$} Ivan Coluzza,\textit{$^{a}$} Lorenzo Rovigatti,\textit{$^{a,d}$} and \\ Peter D.~J.  van Oostrum\textit{$^{e}$} }}\vspace{0.5cm}

\noindent \normalsize{
Limited bonding valence, usually accompanied by well-defined directional interactions and selective bonding mechanisms, is nowadays considered among the key ingredients to create complex structures with tailored properties: even though isotropically interacting units already guarantee access to a vast range of functional materials, anisotropic interactions can provide extra instructions to steer the assembly of specific architectures. The anisotropy of effective interactions gives rise to a wealth of self-assembled structures both in the realm of suitably synthesized nano- and micro-sized building blocks and in nature, where the isotropy of interactions is often a zero-th order description of the complicated reality. In this review, we span a vast range of systems characterized by limited bonding valence, from patchy colloids of new generation to polymer-based functionalized nanoparticles, DNA-based systems and proteins, and describe how the interaction patterns of the single building blocks can be designed to tailor the properties of the target final structures.}
\vspace{0.5cm}
 \end{@twocolumnfalse}]

\footnotetext{\textit{$^{\ast}$~ E-mail:  emanuela.bianchi@univie.ac.at}}
\footnotetext{\textit{$^{a}$~Faculty of Physics, University of Vienna, Boltzmanngasse 5, A-1090 Vienna, Austria.}}
\footnotetext{\textit{$^{b}$~Institute for Theoretical Physics, Technische Universit{\"a}t Wien, Wiedner Hauptstra{\ss}e 8-10, A-1040, Vienna, Austria.}}
\footnotetext{\textit{$^{c}$~Dipartimento di Scienze, Universit{\`a} degli studi Roma Tre, Via della Vasca Navale 84, I-00146 Rome, Italy.}}
\footnotetext{\textit{$^{d}$~Rudolf Peierls Centre for Theoretical Physics, 1 Keble Road, Oxford, OX1 3NP, UK.}}
\footnotetext{\textit{$^{e}$~Department of Nanobiotechnology, Institute for Biologically Inspired Materials, University of Natural Resources and Life Sciences, Muthgasse 11/II, A-1190 Vienna, Austria.}}

\section{Introduction}

In the realm of rational materials design, specific structures at the nano- and micro-scale can be conceived for a vast range of technological applications:  particular attention is for instance devoted to non close-packed architectures, that can act as catalysts, filters, sensors, biomimetic scaffolds or drug delivery devices~\cite{NanoparticleArrays2000,BiosensorVelev,Wang2010}. Responsive open structures can indeed be used in biotechnological applications for tissue engineering, sensing and purification where self-repairing, switchable porosities and the ability to capture drugs are required~\cite{supramol,rubbers}. Additionally, materials containing ordered arrays of holes are also very interesting for functional devices in fields such as photonics, optoelectronics, thermophotovoltaics and for energy storage technologies~\cite{photonicreview}.

Rather than relying on externally-controlled tools, many fabrication methods are nowadays based on self-assembly processes. Self-assembly is essentially the formation of some kind of order, perceived by humans and possibly quantified by some order parameter, as a result of a competition between various enthalpic and entropic factors~\cite{selfassembly2002}. This holds in atomic or (bio)molecular systems as well as on much larger length scales, as for instance in colloidal systems, where the self-assembling units are typically orders of magnitude larger than atoms and molecules. Self-assembly in colloidal model systems is no different than self-assembly in atomic or molecular systems, with the added benefit that the large characteristic sizes allow for the easy inspection of the resulting structures and their dynamics.

In order to gain a greater control over the self-assembly, it is beneficial to impose  additional constraints ~\cite{selfassembly2015}. Extra instructions can be for instance imparted upon the particles if their interactions are no longer merely isotropic but rather depend on their relative positions and orientations~\cite{Zhang_04,Glotz_04}. Present-day approaches to rational materials design try to address the characteristics of the building blocks that are both experimentally feasible and able to stabilize mesoscopic structures of interest, looking meanwhile for the conditions that guarantee the assembly of these units into desired architectures.  In the realm of artificial nano- and micro-sized units, patchy colloids~\cite{Patchy_revExp,patchyrevtheo}, \textit{i.e.}, particles with directional and selective interactions often induced by chemically or physically patterned surfaces, are regarded as promising building blocks for smart materials with designed symmetries and properties. The two main properties of these building blocks, \textit{i.e.}, the limited bonding valence and the directionality induced by the bonding sites, favor those architectures that are compatible with the features of the single unit~\cite{Patchy_revExp,patchyrevtheo}.

Beyond materials science, many naturally-occurring systems are known to give rise to a wealth of self-assembled structures by virtue of some anisotropy~\cite{McManus2016}. The effective interactions between the constituent entities of these systems are often complex and it remains challenging to identify the key elements for guiding and controlling their self-assembling processes. Assembly of biomolecules into supramolecular complexes is at the heart of many biological processes and the dynamic interplay of the different components leads to biological functionality~\cite{Fersht1999}. Complex interactions between biomolecules, such as lysozymes and other proteins, or supramolecular complexes, such as viruses, have been recently described as patchy, meaning that the effective interactions in these systems are characterized by limited valence and directionality~\cite{Lomakin1999,Hloucha2001,Ggelein2008,Fusco2013,Fusco2014,quinn_patchy_proteins,Audus2016,Li2016b}. 

In this review, we present a selection of recent advancements in the field of anisotropically interacting units, focusing on patchy colloidal systems, recently developed soft functionalized nanoparticles and biomolecules. In order to span such a broad range of systems in a concise and clear fashion, we propose a classification of limited valence units according to the path leading to the production or formation of these different anisotropically interacting entities: we refer to patchy particles resulting from either the {\it top-down} or the {\it bottom-up} route. Of course, it is difficult to draw a hard line between the two paths since the emergence of anisotropic interaction is arguably based on self-assembly mechanisms in most of the cases. Here, we make the distinction according to whether or not the resulting units are characterized by internal degrees of freedom: the {\it top-down} route results essentially into hard particles with a pre-defined and fixed patchiness, while the {\it bottom-up} path leads to soft units with flexible bonding patterns. Within the top-down route, many synthesis techniques have been developed to produce functionalized units, mostly allowing the fabrication of patchy colloids where the patch size, number and distribution are determined at the synthesis level and maintained unaltered during the consecutive assembly~\cite{Patchy_revExp,Wang2012colloids,Yi2013,Jiang2009}. In contrast, within the bottom-up path, anisotropically interacting units arise through self-assembly of smaller, possibly flexibly-linked, subunits~\cite{Marson2015,seeman_dna_review_2003}. The particular features of such self-assembled patchy aggregates are the extremely soft interactions and the possible fluctuations in the number, positions and/or in the size of the patches. Not all routes towards functionalized units can be classified in a neat way:  for instance, patchy particles obtained by grafting polymer brushes to the surface of colloidal particles~\cite{Kumacheva_2016} are the result of a self-organization process, but those patchy units do not have the degrees of freedom associated to a flexible patchiness. An other example where the proposed classification becomes  ambiguous is represented by a new type of polymeric networks named vitrimers:  vitrimers are malleable materials that can rearrange their topology without changing their average connectivity~\cite{leibler_science}; the phase behavior of these systems has been recently described with a patchy model that combines a non-flexible particle geometry with a bond switching dynamics that mimics the bond exchange mechanism at the basis of the extraordinary properties of vitrimes~\cite{frank_vitrimers}. 
In addition, the proposed classification is inherently fluid as some systems might be categorized differently depending on the specific point of view: proteins can, for instance, be described as (charged) ``top-down" patchy colloids or as ``bottom-up" complex units (emerging from the folding process) with directional interaction sites. Nonetheless, with all its limitations, the proposed classification provides  insight in the interplay between the design of anisotropic interaction patterns and the balance between entropy and enthalpy during assembly.

This review is organized as follows: in section~\ref{sec:top-down} we present an overview of the experimental advancements in the field of top-down patchy colloids together with the related theoretical/numerical progress on the understanding of their self-assembly, while in section~\ref{sec:bottom-up} we focus on bottom-up patchy units, again presenting both the experimental and theoretical/numerical state of the art. In both sections, we propose a selection of systems that we consider particularly promising to obtain a greater control over self-assembly processes. In particular, the discussion on top-down patchy systems focuses on the following topics: in subsection~\ref{sec:ipcs} we discuss results accumulated so far on charged particles with oppositely charged patches, in subsection~\ref{sec:patchyproteins} we propose a brief discussion on globular proteins in solutions that have been recently described as patchy entities, in subsection~\ref{sec:patchy-polyhedra} we present the interesting interplay between non-spherical shapes and anisotropic bonding patterns, while  in subsection~\ref{sec:patchy-polymers} we touch upon the tantalizing perspective offered by encoding instruction for self-assembly in flexibly-linked sequences of different types of patchy particles. Within the discussion on bottom-up patchy systems we focus on the following topics in more detail: the spontaneous formation of patchiness in polymer-based systems, reported in subsection~\ref{sec:star-polymers}, the effect of anisotropic interactions in the hierarchical self-assembly of DNA-based systems, discussed in subsection~\ref{sec:DNA}, and the relevance of patchiness in biological systems, analyzed in subsection~\ref{sec:proteins}. Finally, in section~\ref{sec:conclusions}, we draw our concluding remarks. 

\section{Patchy particles from the top-down route}\label{sec:top-down}
The common characteristic of anisotropically interacting units resulting from top-down synthesis approaches is the negligibility of fluctuations in the particles surface pattern and/or shape. In other words, most of the instructions for the self-assembly of these units is imparted at the synthesis level either by selectively modifying the surface of the particles or by following experimental protocols leading to non-spherical shapes, while any fluctuations are frozen in as permanent polydispersity. What is left to tune the self-assembly with after synthesis is the physical and chemical properties of the dispersing medium and the presence or absence of external fields.

The simplest type of patchy colloidal system is represented by the special case of Janus particles~\cite{Lattuada2011,Walther2013}, \textit{i.e.}, particles with two distinguishable hemispheres characterized by different surface properties. Fabrication methods yielding large volumes of Janus particles and guaranteeing a precise control over the properties of the two hemispheres have undergone impressive steps forward. Unfortunately, only a few of the developed techniques can be extended to produce other surface decorations of patchy colloids.
A very helpful classification of synthesis methods of patchy particles was presented in reference~\cite{Patchy_revExp}, but we also encourage the interested reader to consult other more recent experimental overviews on patchy colloids~\cite{Yi2013,Duguet2016}. In the following, we briefly summarize the most used synthesis methods with the aim of providing an idea about present-day experimental challenges, such as the fine control on the surface patterns (size, shape, position and orientation of the patches), the richness of the pattern morphologies (number of patches per particle), and the scalability of the methods (amount of synthesized particles).

A class of surface modification techniques that has been successfully applied to make several types of patchy particles for self-assembly studies is {\it glancing angle deposition} ~\cite{Pawar2008}; here a layer of material, usually a metal or an oxide, is deposited from a glancing angle on particles in an ordered two-dimensional array ~\cite{Pawar2008} or in the grooves of templates~\cite{He2012}, to influence the size and shape of the patches through the casting of shadows. The subsequent etching of deposited patches to make these smaller~\cite{Chen2011Etch} can be combined with the possibility to lift off and flip the entire particle array to proceed with the modification of patches on the other side of the particles~\cite{Pawar2009}. The interactions between the patches can then be determined by a specific surface coating of the deposited material; for instance a gold layer can be functionalized with a thiol to make attractive, hydrophobic patches~\cite{Chen2011Kagome} or charged regions~\cite{Hong2006glad}.
Another important category of synthesis methods is {\it templating}, in which some sacrificial material is used to temporarily shield part of the particle surface during its fractional modification. Most templating techniques yield Janus-like particles, but there are exceptions that yield two patches on spherical particles~\cite{Lin2010,vanOostrum2015} or even non-spherical particles with two polymeric caps resulting from partial etching of electrospun polymeric wires~\cite{Ding2011}.
Finally, a conceptually simple manner to make particles with one or two polar patches is based on {\it contact printing}, usually with a soft polydimethylsiloxane stamp on two-dimensional particle arrays~\cite{Cayre2003}. The size of the produced patches can be varied \textit{via} the stiffness of the stamp~\cite{Jiang2009}, which can be easily modified through the mixing ratio between monomers and cross-linkers, the applied pressure and the amount of ``ink`` used. Both physical interactions such as charge attractions and hydrophobicity~\cite{Cayre2003} as well as chemical bonds can be used to fix the used ink to the particles~\cite{Jiang2009,Tigges2015,Seidel2016}.

To date, the aforementioned techniques allow for a fine control over the patch size and position, while the number of patches per particle is typically limited to one or two and the amount of particles produced in a single batch is little. Great efforts have been devoted to the creation of colloids with rich surface patterns, consisting of many patches arranged in geometrical patterns. In this direction, liquid interfaces and surface tension have been used to provide anisotropy in surprising ways. For instance, particles on the surface of an emulsion droplet form regular aggregates upon evaporation of the droplet~\cite{Manoharan2003}. Controlled patch sizes result upon swelling these regular aggregates with a liquid monomer to be later polymerized~\cite{Wang2012colloids}. Interactions between the patches can be modified and made specific by functionalizing the patch surface with DNA~\cite{Wang2012colloids} or metal-coordination-based recognition units~\cite{Wang2013}. The larger patches made in this way are protrusions on the particle surface. Alternative methods to make particles with protruding patches are based on swelling polymeric particles with additional monomer~\cite{Kraft2011} or on condensing monomer droplets on oxide particles~\cite{Ravaine2012}.  A subsequent polymerisation is then used to render these monomeric patches permanent. A multitude of relatively small protrusions of monomers can be polymerized to form areas with an effective surface roughness. In combination with depletion attractions induced by depletants on the scale of the roughness, neutral patches or protrusions can be realized on otherwise attractive particles~\cite{Kraft2012}. Similarly, dimples left after a two-stage polymerization~\cite{Sacanna2010} or flat faces left after temporary melting at a flat substrate can be used to render inherently isotropic interactions such as depletion~\cite{Ramirez2010,Kraft2012} and van der Waals attractions~\cite{Ramirez2012,Ramirez2013} directional.

Beyond the challenges in the particle synthesis, experimental methods to produce self-assembling building blocks with specific surface patterns also deal  with challenges related to the achievement, and possibly the observation, of the desired assembly behavior. First of all, to facilitate the microscopic study of any self-assembled structure, it is advantageous that the particles are dispersed in a refractive index matching solvent, \textit{i.e.}, a solvent with the same refractive index as that of the particles~\cite{VanBlaaderen1992}. Moreover,  for self-assembly to take place in the bulk, a relatively large amount of particles is needed that should also be density matched to suppress sedimentation or creaming towards a hard wall. This can be achieved by working with small particles: the gravitational length has to be large enough to make sedimentation irrelevant on the timescale of the experiment. However, the resolution of the used microscopy technique sets the lower bound to the particles size. One can afford to work with larger particles by density matching them in a properly chosen solvent mixture~\cite{Kegel2000,Lu2008}. 
Finally, it is important to note that the speed of diffusion and therefore the rate at which any self-assembly takes place depends strongly on the size of the particles. The choice of the composition of a colloidal system for self-assembly experiments is a compromise between requirements on the time scale, the length scale, the density, the refractive index and all those properties determining the colloidal interactions that should ideally lead to the desired assembly behavior, and all these factors complicate the fabrication of anisotropic particles for studies in three-dimensions more than in two-dimensions.

The spontaneous assembly of patchy colloids has been experimentally observed for instance in the formation of strings~\cite{Cayre2003} and micelles~\cite{Hong2006glad} of Janus particles and  in the site-specific aggregation of finite colloidal clusters~\cite{Wang2012} or in the formation of an extended, two-dimensional crystal, known as kagome lattice~\cite{Chen2011Kagome}. It is worth noting that assembly of patchy particles can be further influenced with external fields, especially in the case of conducting patches on dielectric particles~\cite{Velev2008-1,Velev2008-2,Velev2009,Velev2010,Kret-confinement11}.
Under the influence of electric or magnetic fields, the formation of chains, staggered chains or close and loosely packed two-dimensional crystals can be induced~\cite{CubesInFields2014}. An overview of experiments in which external fields under different orientations are used is given in reference~\cite{Bharti2015}. It is also worth noting that particles with an anisotropic surface chemistry can move in, possibly self-induced, gradients, a motion that bears similarities to the active swimming of some bacteria. Active patchy colloids are beyond the scope of this review and we refer the interested reader to a number of excellent review articles~\cite{Elgeti2015,Cates2015,Zttl2016}.

From the numerical and theoretical point of view, many model systems have been developed during the past years to include patchy directional interactions in the pair potential between colloidal particles. Most of the patchy models proposed in the literature are designed to mimic the behaviour of top-down patchy colloids, as they usually consist of isotropically repulsive spherical units carrying a fixed and small number of attractive interaction regions: patches are typically arranged on the particle surface and their positions are fixed in a pre-defined geometry. The most popular models are the Kern-Frenkel model~\cite{KernModel03}, the sticky spots model~\cite{Bianchi06} and the orientational Lennard-Jones model~\cite{Doye07cryst}. Despite their simplicity, these models have been successfully employed to investigate a huge range of phenomena, from the formation of quasicrystalline structures in the absence of an external field~\cite{doi:10.1063/1.3679653,PhysRevLett.110.255503} to reentrant gels~\cite{roldan2013gelling} and reentrant spinodals~\cite{doi:10.1063/1.4974830}, while the self-assembly of such systems has been deeply investigated also under gravity~\cite{PhysRevE.93.030601}, on substrates~\cite{0953-8984-29-1-014001}, or under shear~\cite{C5SM00281H,C6SM00183A}. In parallel, theoretical tools developed to investigate the structure and thermodynamics of top-down patchy colloids have reached a mature stage. In this context, the theoretical cornerstone is provided by the Wertheim theory~\cite{Werth1}, which has been extended to support multi-component mixtures~\cite{C0SM01493A}, patches of different types~\cite{PhysRevE.95.012612,Kalyuzhnyi-nonuniform}, multiple bonds per patch~\cite{doi:10.1063/1.4751480}, ring-forming systems~\cite{C5SM00559K}, systems in confined porous media~\cite{Kalyuzhnyi-confined} and much more~\cite{marshall_micro,chapman_mixture,fantoni2015wertheim}.

The incredible variety of behaviors exhibited by the systems presented above are a testament of the versatility of building blocks with patterned surfaces. In the next sections we present some selected examples of systems composed of top-down patchy particles which we consider particularly promising both to assemble materials with desired properties and to gain a deeper understanding of complex systems that  show a well-defined self-assembly behavior: in section~\ref{sec:ipcs} and~\ref{sec:patchyproteins} we highlight how heterogeneously charged colloids and proteins, respectively, are intrinsically patchy systems, in section~\ref{sec:patchy-polyhedra} we investigate the interplay between non-spherical particle shapes and specific interaction sites on the particle surface, while in section~\ref{sec:patchy-polymers} we describe in detail systems of self-folding colloidal strings composed by patchy units.

\subsection{Charged patchy colloids}\label{sec:ipcs}
Non-homogeneously charged colloids have recently emerged as promising building blocks for target structures with specific properties at the nano- and micro-scale. Colloids characterized by well-defined surface regions carrying different surface charges can be generally described as charged patchy particles; in order to emphasize that these units are a different class of patchy colloids with respect to conventional patchy particles, they are often referred to as inverse patchy colloids (IPCs)~\cite{ipc-review}: the term inverse refers to the fact that, while conventional patchy systems are typically characterized by the presence of attractive regions on the surface of otherwise repulsive particles, IPCs carry extended patches that repel each other and attract those parts of the colloid that are free of patches. This class of systems was originally introduced to describe complex units emerging from the adsorption of charged polyelectrolyte stars onto the surface of oppositely charged colloids~\cite{Blaak_IPC}, but it now includes micro-scale particles manufactured with a ternary distribution of the surface charge~\cite{vanOostrum2015}. Clearly, also Janus particles with differently charged hemispheres~\cite{Cayre2003,Hong2006glad,Sabapathy2016} can be considered as (the simplest case of) IPC units.  It is also worth noting that complex interactions between biomolecules, such as proteins and viruses, have been described as patchy (see also section~\ref{sec:patchyproteins}), making it possible to employ the models and techniques developed for IPC systems to achieve a deeper understanding of the assembly phenomena in these natural systems.

\begin{figure*}[t]
\centering{
\includegraphics[width=\textwidth]{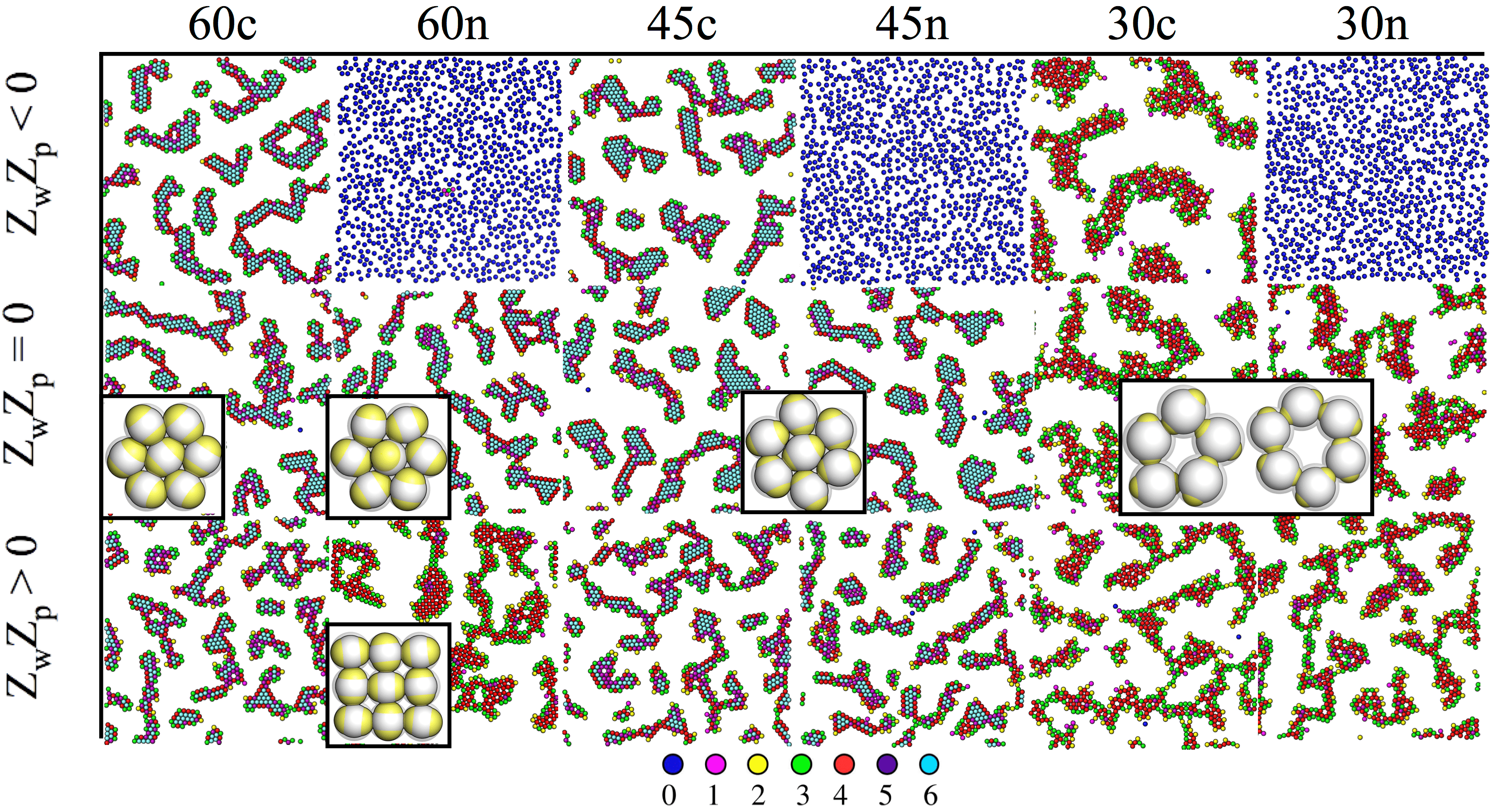}
\caption{IPCs under planar confinement~\cite{bianchi:2d2013,bianchi:2d2014}. Typical simulation snapshots of the several types of IPCs (labelled from the left to right as 60c, 60n, 45c, 45n, 30c and 30n, specifying both the patch opening angle and the net particle charge) under confinement between two parallel walls at distance 1.45 (in units of the particles diameter); the top wall is always neutral, while the bottom wall can be either neutral (panels in the central row, labelled with $Z_{\rm w} Z_{\rm p} = 0$, where $Z_{\rm w}$ is the charge of a unit square on the bottom wall and $Z_{\rm p}$ is the patch charge) or charged (panels in the top and bottom rows, labelled with $Z_{\rm w} Z_{\rm p} < 0$ and $Z_{\rm w} Z_{\rm p} > 0$, respectively). Particles are colored according to the number of bonded interactions; the corresponding color code is displayed at the bottom.}
\label{fig:ipc-confinement}
}
\end{figure*}

Within the framework of materials design, the first elaborate model for heterogeneously charged particles was developed for colloids characterized by two charged polar patches and an oppositely charged equatorial belt~\cite{bianchi:2011}. Subsequently, the idea to study simple models of particles with heterogeneously charged surfaces proliferated within the community. A charged patchy model was for instance developed to study the bulk aggregation behavior of charged Janus-like particles as a function of the patch size~\cite{Cruz2016}. A different modeling of charged Janus colloids was proposed to investigate the relation between the order of the multipolar expansion of the inter-particle potential and the resulting minimum-energy clusters~\cite{Hiero_2016}. In the context of globular proteins, a set of charged patchy particle models was introduced~\cite{Yigit15a} to study the adsorption of such units on a polyelectrolyte chain~\cite{Yigit15b} or a polyelectrolyte brush layer~\cite{Yigit17}. Janus-like dipolar particles and protein-like units have been also modeled to understand how the charge distribution affects the properties of the fluid phase with particular focus on the gas-liquid phase separation~\cite{Blanco2016}. Finally, a highly sophisticated model for particles with icosahedral, octahedral, and tetrahedral charge decorations has been put forward to describe the effective interactions between virus capsids~\cite{Podgornik2013}. 

Within this broad class of systems, we focus in the following on the results accumulated so far for the most studied IPC systems. We refer to the coarse-grained description of the effective interactions between IPCs that was originally developed for particles with two identical polar patches and an oppositely charged equatorial belt~\cite{bianchi:2011} and later generalized to characterize colloids with richer surface patterns~\cite{bianchi:2015}.  The model features hard spherical particles with a surface divided into a few extended regions with different properties; the interaction between two IPCs is characterized  by three independent sets of parameters: the interaction ranges of the different surface areas, their interaction strengths and their surface extents. These parameters can be related to the physical features of the underlying microscopic system -- such as the electrostatic screening of the solvent around the colloids or the charges of the different regions of the particle surface -- by developing a suitable description within the framework of  the Debye-H\"uckel theory for dilute electrolytes~\cite{debyehueckel}. Thanks to this rigorous derivation, the resulting IPC model is not a toy model despite being characterized by a high degree of computational simplicity. 

The distinctive feature of IPC systems is the non-trivial interplay between attractive and repulsive directional interactions: it characterizes the fluid phase~\cite{Yura2015-1,ferrari:jpcm2015} as well as the gas-liquid phase separation~\cite{noya:lamellar2015} and it gives rise to very interesting assembly and phase behaviors. In particular, numerical investigations on IPC systems with two identical polar patches have shown that an emerging feature of such systems is the formation of planar aggregates either as monolayers closed to a charged substrate~\cite{bianchi:2d2013,bianchi:2d2014} or as bulk equilibrium phases~\cite{noya:lamellar2014,noya:lamellar2015}.

Close to a homogeneously charged substrate, IPCs with two identical patches form assemblies with well-defined translational and orientational order depending on the charge ratio of the different entities involved, the patch size and the interaction range (see Figure~\ref{fig:ipc-confinement})~\cite{bianchi:2d2013,bianchi:2d2014}.  
The features of these assemblies depend on the system parameters and were proven to reversibly respond to changing conditions such as the pH of the solution and the charge of the substrate~\cite{bianchi:2d2014}. It is worth noting that the same morphological features observed in simulations have been found in experimental samples of IPCs sedimented on a glass substrate~\cite{vanOostrum2015}; however, while all clusters in the same numerical sample have the same spatial and orientational order, experimental IPCs form different particle arrangements within the same sample, probably due to the patch polydispersity.

In the bulk phase, a tendency towards two-dimensional ordering has been often observed. Three-dimensional structures with two-dimensional order are stabilized by several IPCs systems with two identical patches. For overall slightly charged IPCs with relatively extended and long-ranged patches, a crystal formed by parallel monolayers is stable in a wide region of the temperature versus density plane~\cite{noya:lamellar2014};  the region of stability of this laminar phase depends on the system parameters, \textit{e.g.}, it expands upon increasing the charge imbalance (that is, on changing the pH of the solution) and/or upon reducing the interaction range (\textit{i.e.}, on changing the salt concentration of the colloidal suspension)~\cite{noya:lamellar2015}. The formation of a similar lamellar structure can be also observed in a system of overall neutral IPCs. In contrast to the slightly overcharged case, neutral particles self-assemble into a lamellar phase which confines a disordered phase of mobile particles between the monolayers~\cite{silvanonanoscale}. Lamellar phases represent just one of the many assembly scenarios offered by IPCs: depending on the chosen parameters, IPCs with two identical patches can form an even wider zoo of exotic structures, among which there exist porous bulk phases characterized by parallel nano-channels~\cite{Guenther_thesis_2012}. The control over the properties and the stability range of these phases will open up possibilities to build three-dimensional lattices with an open and very regular architecture, that might be used, \textit{e.g.}, for drug delivery purposes.

\subsection{Patchy proteins}\label{sec:patchyproteins}

It is generally recognized that the interactions between globular proteins are direction dependent~\cite{Lomakin1999,Hloucha2001,Ggelein2008,Fusco2013,Fusco2014,quinn_patchy_proteins,Audus2016,Li2016b}. 
Such a directionality generally stems from the overall shape of the folded protein, the distribution of hydrophobic residues on the protein surface and the distribution of charged residues. At physiological pH values, this results in a heterogeneous surface charge, with the range of the electrostatic interactions being set by the screening provided by ions in solution~\cite{Chan2012,Kurut2012,Blanco2016}. The effect of heterogeneous surface charges and ion condensation on the protein-protein interactions has recently been described in reference~\cite{Lund2016} by an appropriately developed multipolar  coarse-grained model. As pointed out in section~\ref{sec:ipcs}, this class of proteins is related to charged patchy colloids, and several models are being developed in this respect~\cite{Yigit15b,Blanco2016}. The appropriate modeling of directional interactions caused by charge inhomogeneities can be challenging, in particular with respect to specific ion effects: for instance, by forming salt bridges, multivalent ions have very drastic effects on the effective interactions between proteins in solution to the point of rendering directional charge repulsions attractive~\cite{RoosenRunge2014}. More refined models which taking into account electric multipoles as well as the distribution of charged, neutral and hydrophobic residues have been developed\cite{ari2014,McManus2016}.

The phase behavior of dispersions of proteins is of great interest both from a fundamental point of view and for applications. Since the precise folded structure of a given protein is not known \textit{a priori}, detailed information on the shape and surface chemistry is not directly available. However, this information can be obtained through neutron or X-ray scattering experiments with crystallized proteins. Unfortunately, even when the knowledge on the architecture of individual proteins is obtained through scattering experiments after successful crystallization, it is not possible to visualize the self-assembly in a protein solution on the single molecule level~\cite{Li2016}. Consequently, the characterization of the phase behavior of protein solutions is mainly based on interactions with visible light that can be measured with common laboratory equipment and that provide insight on average properties~\cite{Ishimoto1977,George1994,Greenfield2007,quinn_patchy_proteins}. A recent review of the physics of protein self-assembly in the bulk is given in reference~\cite{McManus2016}. In general, the study of the phase behavior of protein solutions relies on many simplifications. It is most common for the shape of proteins to be approximated by spheres, with directional interactions being approximated by a set of patches, the number, distribution, extent and range of which are chosen by mapping average properties measured in experiments. The aforementioned parameters have for instance been chosen such that predictions of second-order thermodynamic perturbation theory matched with experimental information on the critical point~\cite{Ggelein2008} or on the cloud-point (\textit{i.e.} the temperature at which the solution becomes turbid)~\cite{Kastelic2015}. Within these approaches, the range of the interactions has been found to weakly depend on the salt concentration at high ionic strengths~\cite{Ggelein2008}, while the strength of the interactions was observed to depend on the nature of the ions~\cite{Janc2016}.
More complex situations have been recently considered, such as mixtures of different proteins~\cite{Kastelic2016} and non-spherical shapes~\cite{Vcha2011,Kurut2015,Li2015a}.

It is worth noting that the folded structure of bovine serum albumine in solution was recently shown to change in response to variations of the pH and salt concentration~\cite{Sarangapani2015}; a similar conclusion was previously drawn from simulations on the cluster formation in lysozyme solutions~\cite{Chan2012}. It is therefore likely that, at least in some cases, the directional interactions between the proteins depend on the specific conformation in force of the presence of internal degrees of freedom in the proteins. Consequently, these should be considered among the bottom-up systems within the classification proposed in this review (see section~\ref{sec:bottom-up}). On a similar note, it has been recently shown that proteins undergoing amyloid formation change their internal structure during aggregation, providing another clear example of bottom-up self-assembly~\cite{Vcha2014}.

\subsection{Non-spherical patchy colloids}\label{sec:patchy-polyhedra}

Over the last decades, the availability of monodisperse non-spherical particles at the nano- and micro-scale~\cite{Sacanna2011,Vutukuri2014,Miriam2015} has allowed to design a vast variety of structures with different symmetries and packing densities~\cite{DeGraaf2012,Damasceno2012,Schultz2015}. Shape anisotropy can be realized for instance by making use of different growth potentials of crystal planes~\cite{Rossi2011,Miriam2015}, by embedding polymeric particles in a sacrificial polymer matrix that is then deformed~\cite{Champion2007} or even by swelling polymeric particles in a three-dimensional colloidal crystal~\cite{Vutukuri2014}. 
Complex spheroidal shapes range from colloidal molecules~\cite{Manoharan2003,Ravaine2012,Kraft2016} to more complex colloidal aggregates, such as multipod-like clusters of spheres~\cite{Ravaine2016}, while even more anisotropic particles range from convex units, such as rods~\cite{Kuijk2012}, cubes~\cite{Rossi2011} and polyhedral particles~\cite{Vutukuri2014}, to concave shapes, such as tetrapods and octapods~\cite{Miszta2011,Arciniegas2014} or bowl-shaped colloids~\cite{Marechal2010}.
The particle shape plays an important role in the self-assembly of target structures with tailored properties. A systematic study on the assembly behavior of polyhedral hard particles of many different shapes has, for instance, allowed to group polyhedra into four categories of organization: liquid crystals, plastic crystals, crystals, and disordered (glassy) phases~\cite{Damasceno2012}. The same systems can be also mapped according to the coordination number in the fluid phase (\textit{i.e.}, the number of nearest neighbors surrounding each polyhedron in the fluid) and the shape factor (which measures the deviation from a sphere)~\cite{DeGraaf2012}, thus providing a roadmap to drive the assembly into the desired direction. 
At the other side of the spectrum, colloidal branched nanocrystals, like tetrapods or octapods, tend to self-align on a substrate due to their geometry, making them interesting units for technological applications both in two- and three-dimensions~\cite{Miszta2011,Arciniegas2014}. Finally, the particle shape can also be engineered in order to promote lock and key interactions~\cite{Sacanna2013natmat,Wang2014,Ravaine2015}: shapes can be exploited to drive selective lock-and-key binding between colloids, thus realizing a simple mechanism of particle recognition and bonding.

Beyond the already rich framework offered by non-spherical particles, the interplay between the anisotropy of the building blocks and well-defined bonding sites on the particle surface might open tantalizing new perspectives. 

\begin{figure}[t]
\begin{center} 
\includegraphics[width=\columnwidth]{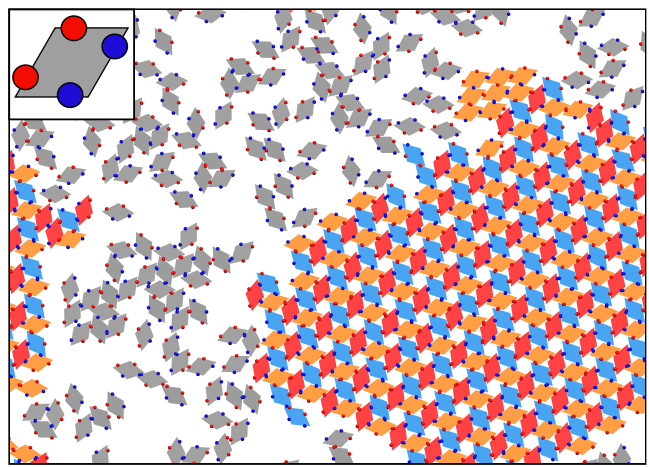}
\end{center} 
\caption{Main panel: self-assembled crystalline monolayer of patchy rhombi organized into an open lattice; the interplay between the particle geometry and the bonding sites controls the resulting phase. The color of the rhombi represents the particle orientation: the unit cell of the lattice is composed of three rhombi with different orientations, represented in orange, red and blue, while the rhombi in the fluid phase are shown in gray; we note that a small crystalline nucleus of a different phase is present in the sample (cluster of orange rhombi on the top-right of the image). Inset: zoom of the single unit.  The color of the patches represents the patch type: two types of patches, A and B, are considered, where only AA and BB bonds are allowed. Patches are asymmetrically distributed along the particle edges.}
\label{fig:patchypolyhedra}
\end{figure}
At the quasi two-dimensional level, the combination of shape- and bond-anisotropy has recently proven to direct the emergence of a rich assembly scenario.  A combined numerical and experimental investigation has shown that unconventional long-range ordered assemblies can be obtained for a class of highly faceted planar nanocrystals only when combining the effect of shape anisotropy with directional bonding~\cite{Ye2013}. Numerical investigations have also shown that regular polygonal nanoplates can be designed to assemble into many different Archimedean tilings~\cite{Millan2014}: the competition between shape anisotropy and interaction patchiness was tuned to obtain either close-packed or open tilings, the latter emerging mainly in binary mixtures of different shapes. Moreover, a systematic numerical study of the assembly of convex hard nanoplates combined attractive edge-to-edge interactions with shape transformations~\cite{Millan2015}: results were divided into space filling (not necessarily regular) tilings, porous (periodic) tilings and complex (disordered) tilings, thus providing an important insight into how shape and attractive interactions can be exploited to target specific tiling architectures. The development of heuristic rules for the design of two-dimensional superlattices would allow experimentalists to improve the crystal properties of already available structures as well as to access exotic phases with new interesting features.   Progress in the synthesis of anisotropic nanoplates offers indeed many possibilities for the rational design of two-dimensional materials with, for instance, different optical and catalytic properties~\cite{Efros2011,Gratzel2011}.  It is worth noting that hard polyhedral tiles decorated by attractive patches can be also used to describe two-dimensional molecular networks~\cite{Blunt2008}: numerical investigations on rhombus tilings have for instance identified the mechanisms leading to the emergence of ordered or random phases~\cite{Whitelam2012}. This study was successively extended by considering model molecules with particular rotational symmetries and studying their self-assembly into network structures equivalent to rhombus tilings~\cite{Whitelam2015}. Along similar lines, the assembly under planar confinement of patchy rhombi with a fixed geometry -- inspired by recently  synthesized particles~\cite{Miriam2015} -- has been investigated~\cite{Carina}, focusing on how the number, the type and the position of the patches along the edges of the rhombi influence the tilings (see Figure~\ref{fig:patchypolyhedra}).

In the bulk, the combination of shape- and patch-induced  directional binding has recently provided amazing examples of tunable ordered structures. 
Binary mixtures of different shapes with mutual attraction induced by isotropically distributed, complementary DNA strands can already exhibit a wide range of exotic extended architectures~\cite{Lu2015,diamond_origami} (see also section~\ref{sec:DNA}): (i) anisotropic polyhedral blocks and spheres, for instance, assemble into complex superlattices that can be tuned by the choice of the DNA shells and the particle size mismatch between the two components of the mixtures~\cite{Lu2015}; (ii) rigid tetrahedral DNA origami cages and spherical nano-particles can form a family of lattices based on the diamond motif~\cite{diamond_origami}.
When focusing on one-component systems of anisotropic particles decorated with anisotropic bonding patterns, most of the results accumulated so far in the literature deal with Janus-like non-spherical entities -- mainly elongated shapes carrying one or at most two patches -- assembling into a vast variety of fiber-like structures with diverse applications. Janus nano-cylinders that form vertical, horizontal or even smectic arrays~\cite{Smith2013,Smith2014}, ellipsoids with one patch in a Janus-like or ``kayak" fashion that form ordered assemblies~\cite{Shah2013} or even field-sensitive colloidal fibers~\cite{Shah2014},``Mickey Mouse''-shaped colloidal molecules that form tubular aggregates~\cite{Kraft2015}, and silica rods coated with gold tips that self-assemble into different multipods~\cite{Chaudhary2012} are just few examples. The susceptibility of Janus-like anisotropic units to external fields can also be used to drive the assembly into string-like structures~\cite{ShieldsIV2013}.

Finally, it is worth noting that the combined approach of complex shapes and bonding surface patches can provide insights into biological processes~\cite{DenOtter2010}. 

\subsection{Patchy polymers}\label{sec:patchy-polymers}
A large proportion of self-assembling processes in living systems adopt a modular approach based on the hierarchical assembly of simple units into larger heterogeneous objects. 
In particular, many biopolymers consist of a rather restricted set of different chemical units, called residues, that, once assembled into linear sequences, selectively acquire specific molecular functions and self-assembling properties. 
The use of a limited set of residues (20 for proteins and only 4 for DNA/RNA) defining a finite alphabet has the advantage that new target structures can be designed (\textit{e.g.}, through evolution) by just changing the order of the elements along the chain. Thus the same alphabet can be used for an efficient recycling of the precious residues: by disassembling chains that do not fulfill their purpose, waste in the form of isolated residues can be efficiently reused for new chains. Incidentally, this is why living organisms can eat each other and synthesize their own proteins from the ingested building blocks. 

Understanding how the linear information is translated into a three dimensional structure is usually referred to as the ``protein folding problem" which will be discussed later in section~\ref{sec:proteins}. Translating the folding property into a purely artificial system would open up new possibilities for the design of novel materials. The assembly of nano- and micro-scale particles into self-folding strings could enable the cost-effective production of responsive meta-materials with unprecedented spatial control over the single particle positions. Indeed, the possible applications of such materials are extremely diverse: from nanoscale switches and sensors that respond to, for instance, temperature, light or pH, to catalysts (mimicking the spatially-defined catalysis of proteins)~\cite{Siegel2010a,Gannavaram2014,Sirin2014} and materials with three-dimensional connectivity that can be used, \textit{e.g.}, to optimize charge separation in photovoltaics~\cite{C6SC01306F,Mejias2016b}.

\begin{figure*}[t]
\begin{center}
\includegraphics[width=\textwidth]{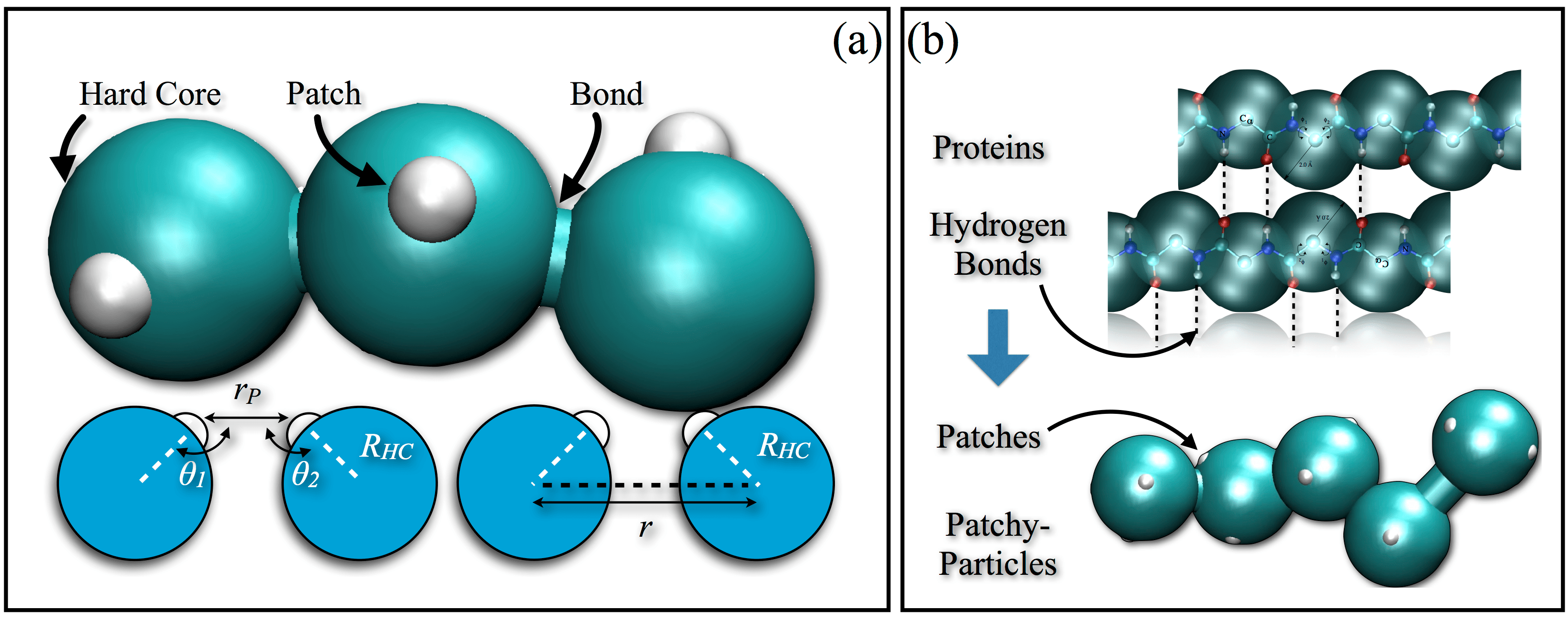}
\end{center} 
\caption{Panel (a). Schematic illustration of a patchy polymer with one patch per particle (top): the isotropically interacting particles are represented by turquoise spheres, while the patches on the colloidal surface are depicted as small white spheres; schematic illustration of the geometrical parameters of the model (bottom): $r_{\rm P}$ is the distance between patches on different chain units, while $\theta_1$ and $\theta_2$ are the alignment angles between these patches (these parameters determine the inter-particle directional interaction),
$R_{HC}$ is the radius of the colloids, while $r$ is the distance between the particles (which solely determines the inter-particle isotropic interaction). Panel (b). Schematic comparison between the protein structure (top) and the Caterpillar model (bottom): both the hydrogen bonds and the patches along the protein backbone pre-sculpt the configurational space, allowing for a successfull design.}
\label{fig:patchypolymer}
\end{figure*}

In order to reproduce the protein folding process at the nano- and micro-scale level, patchy particles can prove themselves essential: anisotropically-interacting units arranged in strings could play the same role as the residues constituting the proteins, effectively constructing functionalized colloidal chains, the so-called ``patchy polymers''~\cite{Coluzza2013,Coluzza2012c,Coluzza2012b} (see panel (a) of Figure~\ref{fig:patchypolymer}). 
Simulations have shown that the control over the folding of particle chains can be obtained combining three main ingredients~\cite{Coluzza2013,Coluzza2012c,Coluzza2012b}: (i) the availability of chains composed of units interacting through different isotropic potentials, (ii)  the control over the particle sequence, and (iii) the presence of directional attractions between the chain units. While the isotropic interactions mimic the chemical flavor of the constituent units along the chain, the directional interactions, induced by, \textit{e.g.}, surface patches, mimic the hydrogen-bonds along the backbone of natural biopolymers. The imprinting of the target structure using a finite alphabet is guaranteed by a reduction of the number of accessible configurations: both the hydrogen-bond and the patch-patch contact networks (in analogy to the secondary structure in proteins) increase the statistical weight of the configurations that are geometrically compatible with the bond directionality. 
The generality and versatility of the self-assembly strategy described above has been demonstrated by the successful design and folding of chains made of just two different isotropically-interacting units, each decorated with two patches~\cite{Coluzza2012c}. Interestingly, knot-like structures can be chosen as targets for the folding and, in analogy with proteins, such knotted configurations can significantly change the melting temperature of these targets upon cyclization~\cite{Coluzza2013}. Further work to characterize the relative designability of the knotted topologies is in progress. Recently, the designability properties of linear patchy polymers have been exhaustively explored in terms of alphabet size and number of patches~\cite{Chiara2017}. An interesting re-entrant behavior emerged, indicating a ``designability window'' for the  number of patches. The boundaries of the re-entrant region are defined by a number of patches  high enough to reduce the number of statistically relevant compact configurations and low enough to preserve the anisotropic nature of the monomer-monomer directional interactions. Such a systematic exploration led to the formulation of a simple criterion to predict \textit{a priori} the designability of different polymer architectures -- a criterion that also natural proteins fulfil. This criterion is based on the appearance of a particular peak in the radial distribution function that dominates over the random packing of the patchy units and is thus experimentally accessible~\cite{Chiara2017}.

The results accumulated so far suggest that designability can be controlled by any strategy that reduces the configurational entropy per monomer, directionality being one example. It would thus be possible to define a general relationship between the configurational entropy per monomer and the alphabet size that controls the boundaries of the designability window, thus serving as a universal guideline to achieve designability. Novel experimentally realizable polymer systems could be conceived using such a guideline.

\section{Patchy particles from the bottom-up route}\label{sec:bottom-up}
All systems discussed in section~\ref{sec:top-down} have in common that the particle bonding pattern is permanent and does not change during the interactions with other units. As pointed out in the previous section, the production of patchy units via top-down approaches has some limitations that might hinder the successful assembly of novel materials. Some of these limitations can be overcome by employing molecular building blocks undergoing a hierarchical self-assembly process: the basic idea of the bottom-up route is to use microscopic constituents, such as polymers or biopolymers, that are able to self-assemble into nano- and micro-scale objects, which, in turn, can generate supramolecular assemblies. Bottom-up approaches allow to obtain precisely defined units, with high yield, precision and monodispersity, once the appropriate sub-units are chosen or designed; the product of the first self-assembly stage, which in the field of proteins is referred to as \textit{folding}, is a finite structure that successively undergoes a second assembly step. The general strategy relies on the spontaneous emergence of the structure of the intermediate constructs, which stems directly from the microscopic properties of the basic molecular constituents: the latter already contains information about the former. In this section, we focus on systems where, at the level of the intermediate constructs, directional bonding and limited valence emerge;  these intermediate objects can thus be referred to as patchy units.  The emergence of patchy-particle-like constructs via hierarchical self-assembly yields particles that are different and inherently more complex than their hard counterparts: being a result of the, often reversible, bonding between smaller objects, these patchy units have many internal degrees of freedom and, as a consequence, are intrinsically floppy. Interestingly, \textit{softness} itself is an emergent trait of this class of particles, and as such can be tuned by changing the microscopic constituents, hence allowing for a systematic study of its role on the dynamics and thermodynamics of soft-matter systems. The presence of inner degrees of freedom in the particles has been proven to play an important role in the determination of the thermodynamics of soft-matter systems~\cite{smallenburg2014erasing}, going as far as controlling the stability of whole phases~\cite{Rovigatti2014a}. In particular, it was shown that patchy systems combining low valence, bond flexibility and soft interactions can stabilize the liquid phase in contrast to the solid one even in the zero-temperature limit, meaning that even at extremely low temperatures the entropic term prevails over the energetic term in the free energy balance of such systems~\cite{smallenburg2014erasing}. These results were found for patchy models apt to describe associating fluids or DNA-coated colloids with limited valence, but also tetravalent DNA nanostars, i.e. extremely flexible patchy units with four bonding patches, were shown to never crystallize and form instead a thermodynamically stable, fully bonded equilibrium gel~\cite{Rovigatti2014a}. 

In recent years, polymer-based macromolecules or nanocomposites (such as mixtures of polymers and colloids) were shown to be extremely powerful and versatile systems for the self-assembly of soft functionalized nanoparticles~\cite{Marson2015}. Janus-like, triblock, striped and multi-patch units can indeed be created by playing with the topology, the geometry and the type of polymeric macromolecules~\cite{Higuchi2008,Cheng2008,Hermans2009,Hanisch2013,Groschel2013,Wang2016}. Compared to their hard counterparts, soft functionalized nanoparticles from polymer-based systems offer a large and easily-accessible playground to direct the design of the desired functionalization. On the other hand, the real time and real space visualization of polymer-based systems through light microscopy is practically impossible because of the small sizes of the forming particles and patches, that are, in turn, limited by the lengths of the polymers. Electron microscopy allows to obtain an insight in the structures present in the sample, but the dynamics are hard to follow and one should be always aware of possible artifacts introduced by the sample preparation.  It should also be noted that, in most cases, the patchiness of the resulting units is frozen in; nonetheless, in principle, these patchy units can be allowed to rearrange by changing the solvent conditions. In section~\ref{sec:star-polymers}, we report several examples of control over the formation of anisotropic units through the specific features of polymeric macromolecules, showing also how the microscopic parameters can influence the emergence of, e.g., gel-, string- or sheet-like structures~\cite{Higuchi2008,Cheng2008,Hermans2009,Hanisch2013,Groschel2013,Wang2016}. In sections~\ref{sec:DNA} and~\ref{sec:proteins}, we extend our discussion from polymers to biopolymers, where the control over the self-assembled functionalized units is achieved by making use of distinct alphabets of molecular bricks. Proteins are biopolymers composed of 20 different types of amino acids, the sequence of which determines the way in which they fold and -- through their natural conformation -- the function they have in nature; DNA molecules are composed of nucleotides of four different types. The interactions between biopolymers such as proteins and DNA are very specific both in direction and in choice of binding partners. The interactions between DNA monomers for instance is dominated by strongly directional Watson-Crick pairing and has been used in many man-made self assembling systems. Subsection~\ref{sec:DNA} presents recent advancements related to man-made DNA-based systems, while in subsection~\ref{sec:proteins}, we discuss the progress on predicting the folding of proteins as well as the use of protein-ligand interactions in detail. 

\subsection{Polymer-based systems}\label{sec:star-polymers}
Polymers are macromolecules consisting of a sequence of monomeric units, each with specific chemical properties, linked by means of covalent bonds. Polymeric macromolecules can either consist of a sequence of identical units - homopolymers - or of sequences of units interacting selectively with the solvent: solvophobic or solvophilic monomers can be either distributed in groups - block copolymers - or evenly  -  heteropolymers - along the chain.
The selective interactions of the monomeric units with the solvent as well as the physical and topological constraints within each polymer give rise to an intramolecular competition between entropic and enthalpic factors: on one hand, solvophobic monomers tend to minimize their exposure to the solvent, thus giving rise to effective enthalpic attractions between themselves, while solvophilic monomers tend to maximize their contact with the solvent; on the other hand, the tethering between the various monomers and the finite volume occupied by each of them give rise to an intramolecular entropic repulsion. The interplay between entropic and enthalpic factors drives intramolecular self-assembling processes.

A first possibility to obtain nanoparticles with effective directional interactions consists in grafting repulsive polymeric units onto attractive nano-scale colloids. Both experiments~\cite{Akcora2009} and simulations~\cite{Bozorgui2013} have shown that these grafted nanoparticles robustly self-assemble into a variety of anisotropic superstructures, typically observed only in systems of anisotropically interacting units. This result, which stems from the deformability of the polymeric corona, holds both for homogeneously and inhomogeneously grafted units and at low and intermediate grafting densities. 
The characterization  of the directional interaction induced by grafting homopolymeric units onto nanoparticles has been theoretically investigated in reference~\cite{Asai2015}. Both homogeneously  and inhomogeneously grafted nanoparticles were shown to qualitatively behave as Janus particles with patches whose size can be predetermined by analytical calculations and appears to be related to the number of grafted homopolymeric chains and to the the size ratio between the radius of gyration of the grafted macromolecules and the radius of the nanoparticle. 
Such an analytical result has been confirmed by extensive simulations performed on colloid functionalized with a low-density  homopolymeric grafting. These particles are shown  to exclusively belong to  the Janus class irrespectively on the number of grafted chains, thus rendering clear that simple homopolymeric units grafted onto a central core might  not be able to give rise to functionalized patches with more than one patch~\cite{Mahynski2015}.  In order to reach such a goal it thus appears to be important to add some chemical complexity to the system.
Similar to permanently grafted nanoparticles, dendrimeric hydrophobic hosts with condensed hydrophilic polymeric guest molecules were also shown to form aggregates whose architecture is driven by anisotropy. Here the directional interactions arise depending on the amount of guest molecules on the hosts which in turn is dynamically controlled via the guest molecule concentration~\cite{Hermans2009}.

A completely different class of patchy units can be obtained by a careful blending of polymeric ingredients only. 
For instance, Janus-like particles can be obtained by mixing two different phase-separating homopolymers in a solution containing also a non-solvent to both polymers. In this case, the Janus fraction is controlled by the different properties of the two polymers, their molecular weights and the mixing ratios of the various components~\cite{Higuchi2008}. The assembly scenario can be enriched to include striped and multi-patch particles by linking the two homopolymers together and mixing the resulting diblock copolymer with one of the two original homopolymers~\cite{Higuchi2008}. 
Collapse of linear copolymers into particles presenting regions with different levels of hydrophobicity and hydrophillicity had been already shown when copolymeric units had been used to mimic protein like structures \cite{Khokhlov, doi:10.1021/acs.jpclett.6b00144}. By exposing linear chains of block copolymers to a solvent that is bad for both parts, it is possible to make the chains collapse into spherical objects. The subsequent phase separation taking place between the two different blocks can be exploited to obtain particles with heterogeneous surfaces\cite{Zhang2011}. The main strength of such an approach is its simplicity, while the size and distribution of the functionalized regions which self-assemble on the collapsed macromolecule are not easily controllable.
By playing with both the geometry of the polymeric macromolecules and the composition of the solution, more complex scenarios can emerge: patchy micelles can spontaneously assemble from a binary mixture of diblock copolymers having a common polymeric section; the patchiness of the system being determined by the solvent quality through the pH~\cite{Cheng2008}. Similar results can be obtained by imposing different topological constraints, for instance by binding three homopolymers together. Changing the way in which the three types of polymers are linked together controls the functionality of the system: triblock copolymer chains~\cite{Groschel2013} and stars\cite{Hanisch2013} have been shown to self-assemble into patchy particles that then form string-, sheet-like superstructures or even open lamellar and cylindrical aggregates depending on the concentration of particular ions (that modifies the solubility of one of the arms), the volumes occupied by the different groups and/or the solvent quality~\cite{Hanisch2013,Groschel2013}.

\begin{figure}[h]
\includegraphics[width=9cm]{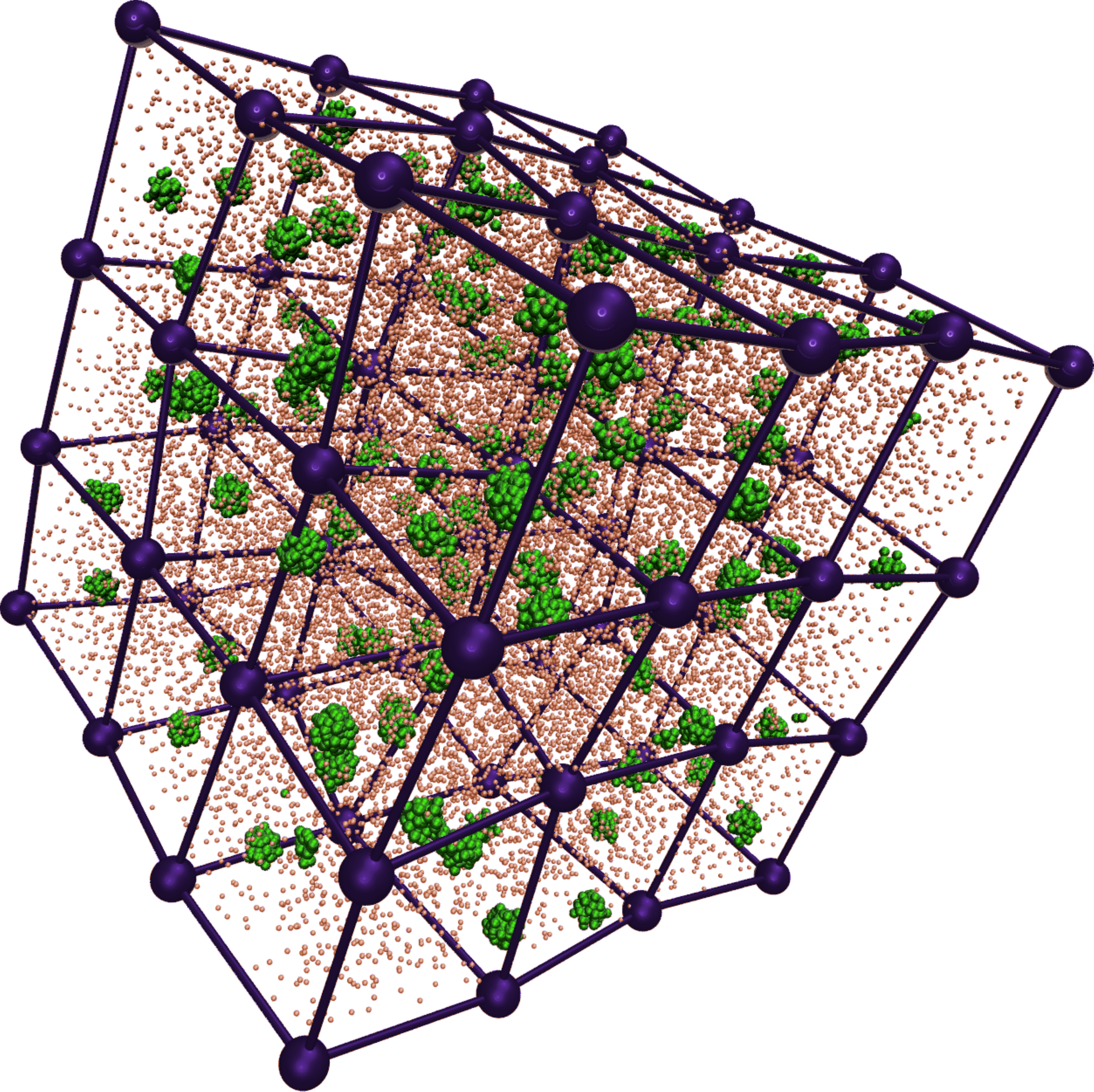}
\caption{Functionalised TSPs in a mechanically stable crystal configuration. TSPs undergo a hierarchical self-assembly process: initially the intramolecular aggregation turns each macromolecule into a soft patchy colloid (here each unit has six patches) which then, in turn, are able to stabilize phases compatible with their functionalization (here a simple cubic crystal).}
\label{fig:cubeTSP}
\end{figure}

Finally, very promising polymeric macromolecules that have been recently shown to self-assemble into soft functionalized units are diblock copolymer stars, also referred to as telechelic star polymers (TSPs)~\cite{tsp-review}. Experimental realizations of such macromolecules are, \textit{e.g.}, end-functionalized star polymers \cite{pitsikalis:mm:1996, vlassopoulos:jcp:1999,Wang2016}. TSPs are obtained by grafting a number $f$ of diblock copolymers, each consisting of a solvophobic and a solvophilic section, onto a central anchoring point:  the solvophilic heads form the soft particle core, while the solvophobic tails are exposed to the solvent and form a shell that can either be equally distributed on the surface or assembled into distinct regions on the surface. In selective solvents, the intra-star association can lead to the formation of soft patchy units. The number and the size of the resulting patches depend on $f$, the percentage of attractive monomers in the polymer chains, $\alpha$, and the solvent properties (temperature, pH, \textit{etc.})~\cite{LoVerso:2006,capone:prl:2012,capone:2013p095002,Rovigatti:2016p3288}. In particular, by exploring the single molecule phase diagram as a function of $f$ and $\alpha$, slightly below the $\Theta$-temperature of the solvophobic part, it was shown that systems self-aggregate into patchy assemblies. The impressive peculiarity of TSP systems is that the stars retain at finite density the valence and the patchiness observed at zero density. Thanks to their robust and flexible architecture, TSPs have been dubbed ``soft Legos''  that possess the ability to self-assemble at different levels: at the single-molecule level, TSPs order as soft patchy colloids which then, at the supramolecular level, stabilize complex phases that are compatible with the functionalization of the single, self-assembled units. These supramolecular structures include gel-like networks~\cite{capone:2013p095002} and complex crystal structures~\cite{capone:prl:2012}, such as diamond or cubic phases (see Figure~\ref{fig:cubeTSP}).
It is worth stressing that the number of functionalized (patchy) regions is completely controlled by the choice of a pre-determined set of chemical and/or physical parameters. Indeed, in addition to the permanent parameters ($f$,$\alpha$), the valence of the self-assembled functionalized units can be altered via the solvation conditions, \textit{i. e.}, the chemical composition of the solvent, the pH or temperature~\cite{Rovigatti:2016p3288}, thus rendering TSPs fully versatile and tunable flexible self-assembling building blocks.
The possibility to tune the number and the equilibrium arrangement of the patches, in combination with the capability of these patchy assemblies to maintain their internal structure at finite density, strongly motivated the development of a coarse-grained model inspired by TSPs and well-suited for the investigation of the bulk behavior of these systems. In particular, the model has been used to elucidate the role of softness and patch rearrangement on the formation of gel networks at a level where both features can be controlled by appropriately chosen parameters~\cite{soft_patchy_bianchi}. 

\subsection{DNA-based systems}\label{sec:DNA}
Biological systems use DNA to store and retrieve genetic information. The information is encoded in a linear 
fashion by a four-letter alphabet thanks to the high selectivity provided by the Watson-Crick base pairing. This high 
selectivity, together with the very different mechanical properties of single- and double-strands, makes DNA an 
excellent candidate for bottom-up self-assembly~\cite{seeman_dna_review_2003}. 
Using DNA has a few additional advantages. First of all, nature provides a whole set of enzymes that make it easier 
to work with nucleic acids. Secondly, the cost of synthetic and viral DNA has dropped significantly in 
the last decade. Lastly, DNA- or RNA-based structures can also be employed as medical tools in virtue of
their compatibility with biological matter~\cite{science_nanorobot}. As a consequence, the use of DNA in materials science and nanotechnology has grown considerably in the last few years.

The notion of using DNA as a building block dates back to the 1980s, when N. C. Seeman started working on DNA-based 
materials~\cite{seeman_junctions}. On one side, his seminal work spurred the fast development of what is today 
known as DNA nanotechnology~\cite{seeman_dna_review_2010}, which employs DNA as a tool to build molecular 
motors~\cite{wollman2013transport}, logic gates~\cite{winfree_gates} and finite-sized objects with pre-designed shapes 
such as polyhedra~\cite{mao_polyhedra,winfree_crystals}, tubes~\cite{liu_tubes,yin_tubes} or even complicated, irregular 
structures, \textit{e.g.}, DNA origami~\cite{rothemund_origami}, which can be used as nano-scaffolds for high precision 
experiments~\cite{origami_scaffold_review,tsukanov_origami,force_clamp_origami} or even as drug delivery vectors~\cite{science_nanorobot}. 
On the materials science side, the early results obtained by Seeman and others showed that DNA can be also used as 
a building block for the generation of ordered and disordered bulk phases~\cite{mirkin_96,seeman_dna_review_2003,Jones2015,seeman_3d_crystal}. 

DNA-based soft-matter building blocks have been historically divided into two distinct categories: DNA-coated colloids 
(DNA-CC) and all-DNA supramolecular constructs (all-DNA). Both strategies exploit the selective binding provided by the 
Watson-Crick mechanism to introduce an effective, temperature-dependent interaction between the basic constituents of 
the system, although the two schemes are essentially different and have different advantages and disadvantages. However,
there exist few systems which sit in the middle and can be seen as intermediate between DNA-CC and all-DNA systems. 
Relevant examples are very small particles grafted with very sparse, ultrashort strands~\cite{starr_langmuir} and mixtures
of DNA-CCs and DNA origami~\cite{diamond_origami}.

\begin{figure*}[t]
\centering{
\includegraphics[width=0.8\textwidth]{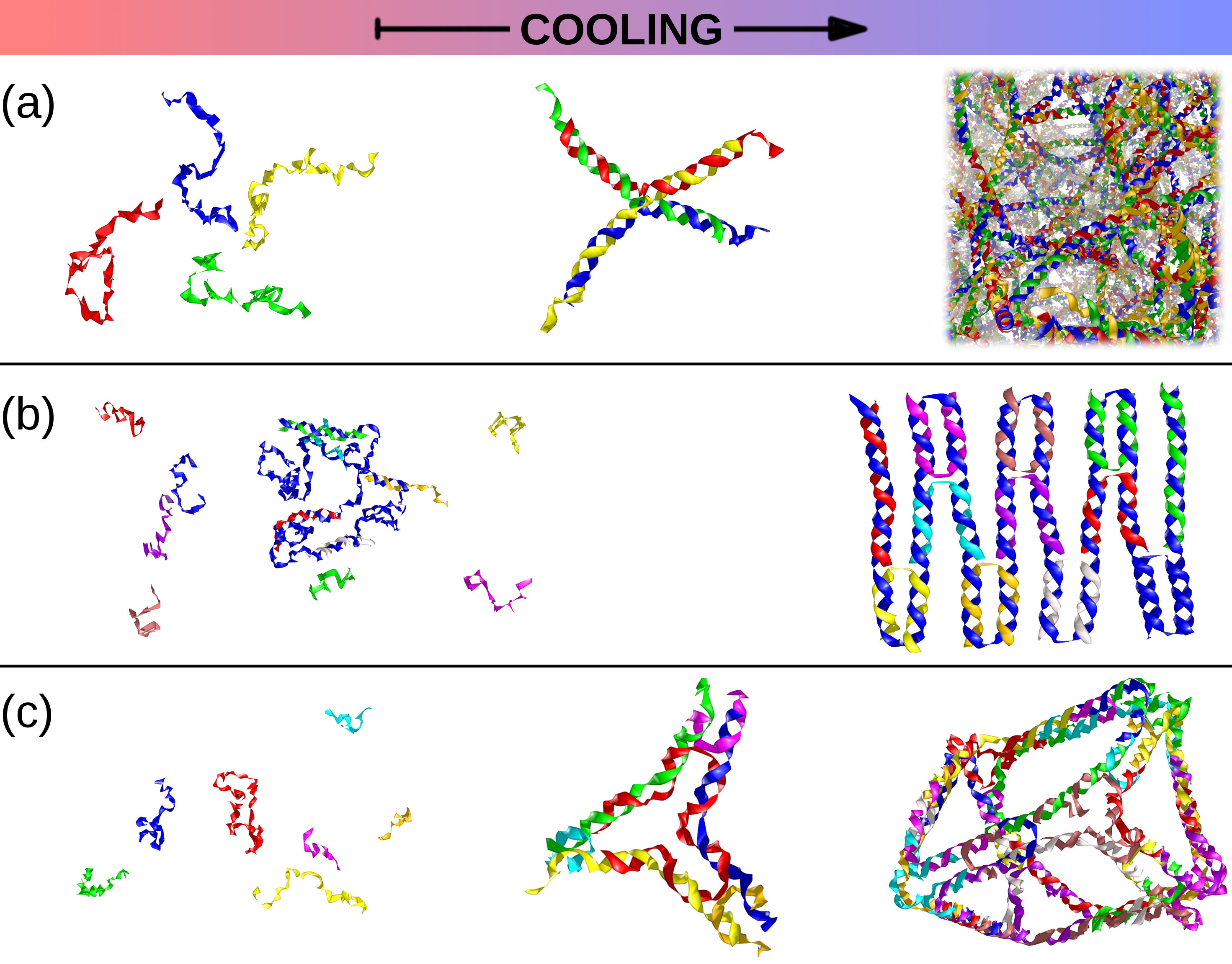}
\caption{\label{fig:hierarchical_dna}Snapshots taken from oxDNA simulations~\cite{doye2013coarse} of all-DNA self-assembling systems. As the temperature, $T$, goes down (from left to right), single 
strands join to form larger structures. (a) The hierarchical self-assembly of an all-DNA gel: as $T$ decreases, single 
strands assemble into tetravalent DNA-constructs which, in turn, link with each other to form a physical 
gel~\cite{biffi_dna,Rovigatti2014}. (b) A solution of a long scaffold strand (in blue) and short staple strands at high 
temperature turns into a small DNA origami as $T$ is lowered (courtesy of Ben Snodin). (c) All-DNA tiles can be used 
as building blocks to assemble more complicated supramolecular structures such as polyhedra or cage-like structures
(courtesy of John Schreck)~\cite{mao_polyhedra,winfree_crystals}.}
}
\end{figure*}

\subsubsection{DNA-coated colloids. }

The possibility to tune the mutual interaction between nano- and microsized particles by functionalising their 
surface makes them suitable for many technological and medical applications~\cite{dellago_dna,eiser_review}. For 
example, polymers have been used for decades to sterically stabilize colloidal solutions~\cite{napper1983polymeric}. 
In addition to providing a tunable repulsion, grafted DNA can also be used to add a controllable mutual 
\textit{attraction} between colloids. By 
carefully choosing the strand sequences and grafting density, DNA-CC have been used to create non-compact 
crystals~\cite{mirkin_96,nykypanchuk2008dna}, crystals with tunable lattice parameters~\cite{maye2010switching}, 
gels~\cite{varrato_bicontinuous_gels} and more~\cite{dellago_dna,eiser_review}. 
Most of these results have been obtained by using a uniform DNA-grafting density. There are a few notable exceptions.
Feng \textit{et al.}, for example, have developed a simple method to make micron-sized DNA-CC with a single patch
and a very high yield~\cite{DNA_patchy_particles}. By contrast, Wang \textit{et al.}, use a much more versatile,
albeit laborious, technique to create ``colloidal analogues of atoms with valence'' by selectively grafting 
DNA-strands on the protrusions of small clusters of amidinated polysterene microspheres with well-defined 
symmetries~\cite{Wang2012colloids}. A similar strategy, which employs DNA-grafted polyhedral blocks to add 
directional binding, has been recently proposed~\cite{Lu2015}.

However, valency can also be enforced in systems made of uniformly-grafted DNA-CC, provided that the DNA strands
can diffuse over the surface~\cite{leunissen_mobile_dna}, as suggested by Angioletti-Uberti 
\textit{et al.}~\cite{Angioletti-Uberti2014}. The basic idea revolves around grafting additional non-binding DNA strands onto the particle surface: thanks to the many-body nature of the interaction between DNA- CC, the resulting particle valence is controlled by the interplay between the non-specific repulsion, which depends on the strand length, temperature and salt concentration, and the attraction due to DNA hybridisation. 
Coarse-grained simulations show that, in contrast with DNA-CC with immobile DNA linkers, these limited-valence DNA-CC can self-assemble into open structures~\cite{Angioletti-Uberti2014}.

The possibility of going beyond uniformly-grafted DNA-CC has been recently demonstrated by the development of techniques aimed at grafting single DNA-strands with specific arrangements on the surface of colloids~\cite{kim_angew,suzuki_jacs}. As shown by Halverson and Tkachenko, these directionally functionalized DNA-CC could provide a route to generate error-free, mesoscopic structures with high yield~\cite{mesoscopic_dna}. 

\subsubsection{All-DNA constructs. }

In the context of anisotropic interactions, and specifically of patchy systems, all-DNA constructs have the 
advantage of being intrinsically valence-limited, since the maximum number of possible bonds can be readily selected
by a careful design of the DNA sequences~\cite{biffi_dna}. However, these constructs require an accurate synthesis and
long purification and annealing protocols~\cite{biffi_dna}. Generally, all-DNA materials require a multi-step 
(hierarchical) self-assembly process: first, single strands assemble into DNA-constructs or tiles. These constructs, 
in turn, self-assemble into larger objects or structures upon lowering the temperature~\cite{seeman_junctions}. 
Figure~\ref{fig:hierarchical_dna} presents a few examples of all-DNA systems undergoing hierarchical self-assembly. This 
strategy has been employed in the past to generate crystalline structures of different types~\cite{seeman_junctions,winfree_crystals}. 
More recently, DNA has been incorporated as a tool to investigate and generate soft-matter disordered or partially ordered
materials. For example, Bellini and co-workers have shown that ultrashort DNA strands can pile up and self-assemble 
into long chains that, at high concentrations, form liquid crystalline states~\cite{BelliniScience07}. The dependence on 
the sequence, the type of nucleic acid (DNA or RNA) and the effect of the presence of dangling ends have all been investigated
experimentally~\cite{BelliniScience07,BelliniJACS08,BelliniJPCM08}, showing that the final state is deeply affected by even 
small changes of the building blocks. A fundamental understanding of the self-assembly of these systems has been 
provided by the theory developed by de Michele and co-workers. Their theory, together with accompanying 
numerical simulations, has shown that the formation of these all-DNA liquid crystals can be understood in the framework 
of patchy particles by modeling double strands as cylinders with two patches, resulting in semi-quantitative
agreement with experiments~\cite{ourMacromol,DeMichele2012}. 

This connection between simple anisotropic toy models and DNA constructs 
has been further strengthened by recent experiments on trivalent and tetravalent DNA nanostars, \textit{i.e.} DNA constructs 
with a fixed valence of three and four, respectively. Bellini and co-workers have carried out measurements of the 
low-density phase diagram and dynamics of these nanostars, providing the first experimental confirmation of the 
dependence of the size of the gas-liquid instability region on the valence~\cite{biffi_dna} and a thorough 
characterization of the structural and dynamical properties of the equilibrium gel phase~\cite{biffi_soft_matter,francesca_epje,jcp_javi}. On the 
numerical side, accurate coarse-grained simulations and liquid-state theories have been shown to match experimental results~\cite{Rovigatti2014,acs_nano_DNA}, supporting the experimentally observed thermodynamic stability of the disordered gel with respect to 
crystallization~\cite{Rovigatti2014a}, in line with recent results on toy models of patchy particles~\cite{frank_nat_phys}. 
In a similar fashion, building on earlier numerical and theoretical work done on toy models~\cite{sandalo_gel}, mixtures of different DNA constructs have been used to synthesize a material that gels on heating~\cite{reentrant_gels}.

The collection of these results suggests that the phenomenology observed in patchy systems can be reproduced by all-DNA systems, provided 
that the intrinsic flexibility of DNA is taken into account~\cite{frank_nat_phys,Rovigatti2014a}. 

\subsection{Biopolymers and proteins}\label{sec:proteins}
Proteins are heteropolymers of different length composed of 20 different types of amino acids. The tasks that proteins perform are very diverse: they usually involve interactions with other proteins or other biomolecules, such as DNA or RNA. These interactions are controlled by the same elements that encode for the native structure of the protein itself. Structure and function are, as a consequence, strongly correlated, and directly dependent on the sequence of amino acids along the protein chain. Hence, unveiling the fundamental properties of proteins is of paramount importance to understand and control the physiology of living organisms at the molecular level. Depending on the amino-acid sequence, some proteins can collapse to form a well-defined ``native'' conformation, while others can not. The process of forming a compact, native structure is referred to as folding. The majority of biochemical reactions in living organisms is based on the activity of proteins~\cite{Fersht1999}. Each protein performs a specific set of functions that generally requires folding the chain into specific configurations \cite{ANFINSEN1973,Fersht1999} or even fluctuating within an ensemble of configurations~\cite{ISI:000178603100013,Fink2005,Dyson2005,ISI:000277219300001}. 
This remarkable property can be exploited to construct novel materials either by creating protein-like systems (\textit{e.g.}, the patchy polymers described in section~\ref{sec:patchy-polymers}) or by creating artificial proteins designed to self-assemble into target architectures. Despite encouraging results, computational protein folding~\cite{Dobson1998,Fersht2002,Karplus2005,Lindorff-Larsen2011,Das2008,Fersht2002,Karplus2005,Dobson1998,Mirny2001,Onuchic2004,Tozzini2005a} and protein design~\cite{Bradley2005,Dagliyan2013,Dahiyat1997,Dahiyat1997a,Desjarlais1999,Desjarlais1995,Hellinga1991,Koga2012,Kuhlman2004,
Mandell2009,Smadbeck2014,Thomson2014,King2012,King2014,Huang2014a} remains a daunting task. 
The potential applications can be divided into three principal categories that could profit enormously from computational protein design: chemistry (\textit{e.g.}, enzyme design~\cite{Eiben2012,Baker2010,Siegel2010a,Jiang2008}), materials science~\cite{Mejias2016,King2014,King2012} and medicine, where computational approaches could significantly contribute to drug design by, for instance, offering new and more powerful delivery methods for existing anti-cancer drugs~\citep{Wang2010,Marelli2013}.

\subsubsection{Super-selective drug delivery vectors. }

Nanomedicine, which uses nanosized, appropriately designed drug delivery vehicles to improve targeting of tumors, is one of the most promising fields for cancer imaging and treatment~\citep{Wang2010}. Nanoparticle vectors typically comprise two major elements: a container including the drug and a smart surface capable of releasing the drug only in the vicinity of the target. The synthesis of such particles poses several problems, from the efficient loading of the drug to the coating strategy of the surface of the particle~\citep{Marelli2013}. 

One of the major challenges to design efficient drug delivery vectors is the engineering of the particle functionalization to discriminate between cancerous and healthy tissues. Being present in high concentrations on different typologies of tumor cell membranes, several biomarkers or receptors~\citep{Evans2004,Petros2004,Willingham2012} would be optimal targets, were it not that they are also present on healthy cells, albeit at lower concentrations. One strategy to overcome this problem relies on multivalency:  by associating  the same drug delivery vector with many ligands, the effective binding to a cell membrane becomes very sensitive to the concentration of the biomarker. Super-selective multivalent drugs bind preferentially to surfaces rich in the target receptors. Multivalent particles are proving to be a promising new direction  in nanomedicine~\cite{Interactions2007,ISI:000076680700004,Mammen1998,ISI:000186205300017,ISI:000222531900035,ISI:000225782300008}, and are in particular  very effective against several types of cancerous cells~\cite{ISI:000076680700004,Mammen1998,ISI:000186205300017,ISI:000225782300008,Interactions2007,Davis2010,ISI:000311815300047,ISI:000320328700076,ISI:000341477600005,ISI:000341409800001,ISI:000331517000021,ISI:000364074300001,Falvo2016}. 
A key point is that the interaction between the nanoparticle ligands and the target receptors must be both selective and weak, so that, in analogy to Velcro\textsuperscript{\textregistered}, the particle only bind to surfaces that have a high concentration of receptors~\cite{ISI:000339091600012,Martinez-Veracoechea2011}. Currently, coatings targeting cancers are identified mainly through large trial and error screenings with limited help from computational modeling, and without specific design to control the binding affinity. The key to the realization of velcro-like particles is in the design of the binding strength and specificity of the ligand-receptor bonds~\cite{ISI:000339091600012,Interactions2007} as well as the control over the geometrical distribution of the ligands on the surface of the nanoparticles. Proteins are an optimal choice as ligands since they offer excellent control over the binding selectivity and can be chosen so as to minimize the interference with normal cell function. Unfortunately, among natural proteins it is very hard to select for ligands with low binding strength while keeping high selectivity. A computational approach might thus provide such control and could be used to design and test new efficient cancer-targeting drug delivery vectors. A possible strategy to achieve the design of multivalent drug delivery vectors will require a protein design method and a way to control the binding affinity without affecting the specificity towards the target receptors.

\subsubsection{Protein design.}

The design of proteins is an example of the so-called ``inverse folding problems'' (IFPs). IFPs consist of the search for amino acid sequences whose lowest free energy state (\textit{i.e.}, the native structure) coincides with a given target conformation. The design of natural protein structures holds its rank in the \textit{hall-of-fame} of the greatest scientific challenges~\cite{Coluzza2017}. 

Full-atomistic protein models are a very successful approach to the computational modeling of proteins. Such models are detailed representations where all atom-atom interactions, including interactions with the solvent molecules, are explicitly taken into account and are used to infer the equilibrium properties  as well as to characterize the protein dynamics.  Examples that summarize the current possibility of computer-aided protein design are: the mutation of buried core residues to alter the inner packing of the target proteins~\citep{Dahiyat1997,Desjarlais1995,Desjarlais1999,Hellinga1991,DiStasio2012} or to increase protein thermal stability~\citep{Kuhlman2004}, the introduction of residues in order to catalyze  specific chemical reactions~\citep{Rothlisberger2008}, the synthesis of entirely novel sequences for existing protein structures~\citep{Dahiyat1997a}, and finally the design of proteins not found in nature~\citep{Degrado1999,Betz1993,Chino2015,Whitehead2012,King2012,Kuhlman2003}.
The major drawback of full-atomistic models is that they require massive computational resources, which is the biggest limitation to computational drug design. Moreover, none of the methods developed so far has successfully designed proteins with tuned ligand-receptor interaction strengths.

A direct approach to solving the IFPs would be to use large computer facilities to exhaustively screen sequences capable of folding into a target structure. The enormous number of possible sequences that one would need to test is beyond the available computational power (\textit{e.g.}, to a short peptide of just 20 amino acids correspond $\sim 10^{18}$ sequences, out of which only a negligible small fraction would fold into the target structure). Shakhnovich et al. ~\cite{Shakhnovich1994,Gutin1993} developed a protein design strategy based on the Random Energy Model (REM)~\cite{Derrida1981b,Pande2000}, which describes the freezing transition of a heteropolymer in a mean field approach. The validity of the REM solution to the IFPs has been extensively proven using lattice models~\cite{Gutin1993,Shakhnovich1993a,Shakhnovich1993c,Coluzza2003,Coluzza2004,Coluzza2007a,Rubenstein2012,Abeln2008}. Recently, Coluzza~\cite{Coluzza2011,Coluzza2014} introduced the Caterpillar coarse-graining model based exactly on the idea that the proteins are pre-sculpted by molecular features, thus extending the REM approach off-lattice. In the Caterpillar protein model, the amino acids are represented as overlapping spheres anchored along the backbone (see panel (b) of Figure~\ref{fig:patchypolymer}). The model has only transferable parameters that have been adjusted so that the designed sequences refolded to their native structures~\cite{Coluzza2014}. 

\subsubsection{Super-selective ligand design. }
\begin{figure}[h!]
\begin{center}
\includegraphics[width=\columnwidth]{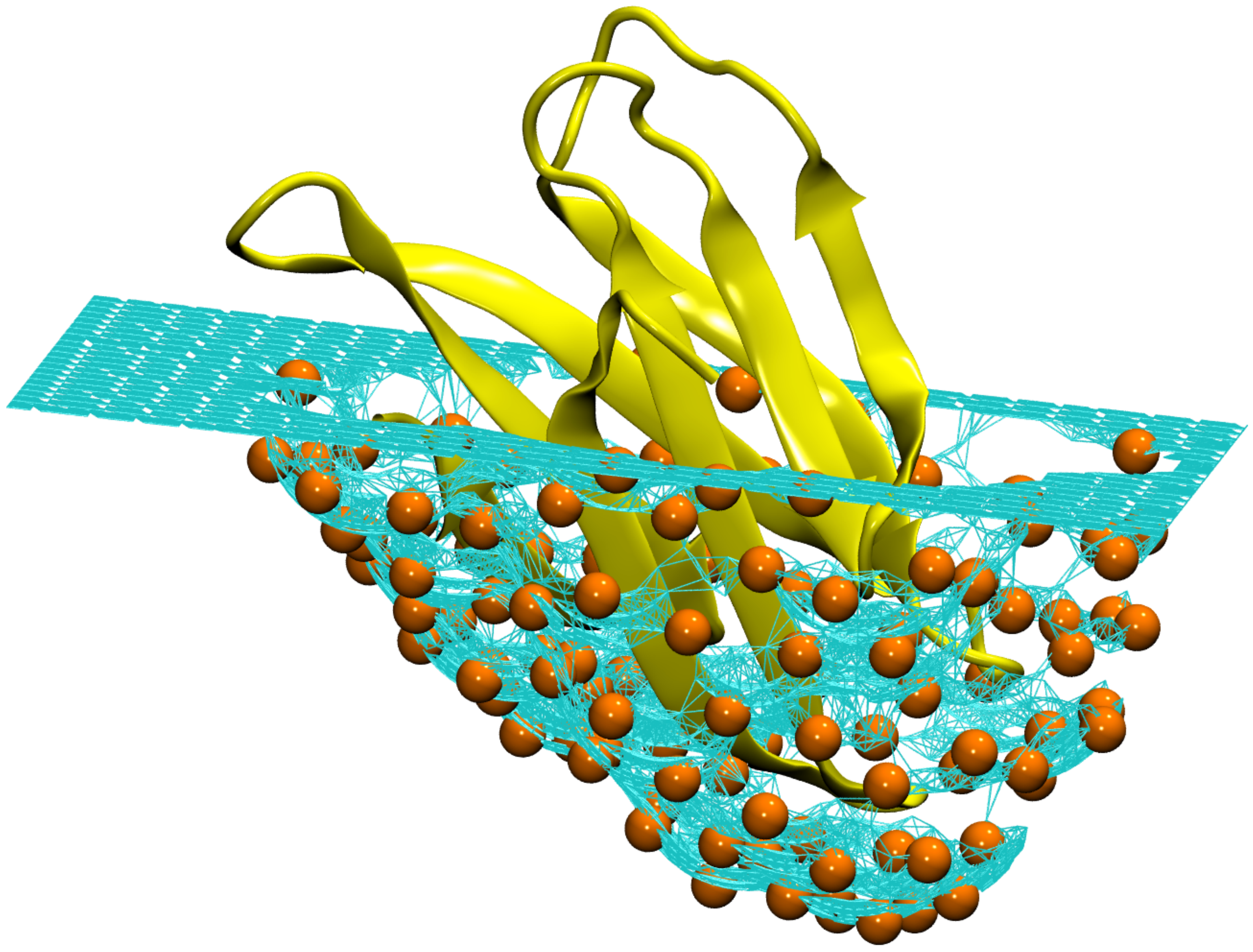}
\end{center}
\caption{Schematic representation of an ideal pocket designed to reproduce the structure of the human Fas apoptotic inhibitory protein (PDB Id. 3mx7)~\cite{Nerattini}. The turquoise surface covering the protein (colored in yellow) is made of a mesh of hard particles (colored in orange) with the radius equal to the hard core radius of the Caterpillar model ($2$\AA). When the protein is pushed into the mesh, the hard particles are repelled by the self-avoiding interaction. Different surfaces can be obtained by changing the orientation of the protein and the maximum depth. It is interesting to notice that this procedure is reminiscent of the experimental techniques used to produce molecular moulds~\cite{El-Sharif2015,Richter2014,Puoci2010,Vasapollo2011}.}
\label{fig:pocket}
\end{figure}

One of the key properties of biological molecules is that they can bind strongly to certain compounds and yet interact only weakly with the very large number of other molecules that they encounter.  Simple lattice models were used to test several methods to design binding specificity~\cite{Coluzza2004}. It was shown that binding sites designed to interact quite strongly with compounds are unlikely to bind non-specifically to other molecules. Such a specificity can be further controlled if binding occurs simultaneously with folding. Lattice proteins that do not fold in solution were observed to undergo interaction-induced folding~\cite{Coluzza2007a}. These proteins bind with the same high specificity as proteins already folded in solution, but have a considerably lower binding free energy. In other words, these proteins can bind to a compound in a way that is highly specific, yet reversible.  On one hand, the specific binding can be achieved by designing the target protein folded and bound to the receptor. On the other hand, to reduce the binding strength, few random residues on the non-binding region of the protein-receptor complex can be forced to be hydrophilic, setting the balance between the folded-bound and the unfolded-unbound state. Although this scheme has extensively been tested only on lattice proteins, a preliminary study using the Caterpillar model seems to confirm that the same working principle applies in the continuum~\cite{Nerattini}. In fact, the preliminary results obtained so far with this model show that it is possible to design proteins so that they bind specifically to pockets tailored to the ligand native structure (see Figure~\ref{fig:pocket}).  The proteins are designed to fold in a pocket and, depending on the pocket design and on the part of the sequence that does not interact with the pocket, the propensity of the protein to also fold in solution can be tuned. The binding affinity of the designed proteins can be controlled in the same way as for lattice proteins, \textit{i.e.}, by using the randomness of the artificial sequence. 

\section{Conclusions}\label{sec:conclusions}

Self-assembly is a fundamental mechanism that rules nature at the atomic/molecular scale and is nowadays exploited in materials science to build desired structures at the colloidal (nano- or micro-) scale. The spontaneous formation of a target equilibrium architecture is profoundly affected by the thermodynamic balance between entropic and energetic contributions as well as by the aggregation kinetics of the self-assembly process. The interplay between these factors can be to some extent controlled when the self-assembling units possess themselves some additional information for their spontaneous organization. The newest and most successful routes to self-assembled materials rely on anisotropy: extra instructions for the assembly of target materials with desired architectures and properties can be imparted upon the particles if the interactions are no longer merely isotropic but rather depend on the relative positions and orientations of the particles with respect to each other. 
In this review, we have focused in particular on soft matter systems characterized by a limitation on the number of bonds that each unit is able to form: by virtue of the asymmetry and selectivity of their interaction patterns, particles with limited bonding valence are excellent base units for the spontaneous assembly of desired structures. We considered a broad range of systems and traced out some of the emerging trends in the design of functionalized particles for the assembly of materials with tailored properties. Contextually, we proposed also some related insights on the mechanisms behind some naturally-occurring phenomena such as protein folding and protein crystallization. The topics touched upon in the present review are highly interdisciplinary, involving physics, chemistry, chemical engineering and bio-related sciences, so that a unique framework is hard to be drawn. Among the rich variety of units characterized by limited bonding valence, such as patchy colloids, proteins, polymers and DNA-based systems, we focused on a selection of functionalized building blocks that we believe to be very promising. In particular, we reported in detail on some ingredients of self-assembling systems that might be crucial to gain an ever greater control over the forming structures, such as emerging directionality between charged units, the role of pre-existing bonds between some of the assembling units, the interplay between anisotropic particle shapes and bonding patterns, enhanced fluctuations of the bonding patterns, and/or the control over the bond specificity. 

\section*{Acknowledgments}
The authors are deeply indebted to C. Dellago, G. Kahl, C.~N. Likos and E. Reimhult for fruitful discussions. EB acknowledges support from the Austrian Science Fund (FWF) under Proj.~Nos. M1170-N16 (Lise Meitner Fellowship) and V249-N27 (Elise Richter Fellowship). BC acknowledges support from the Austrian Academy of Sciences ({\"O}AW) through the Grant No.\ 11723. IC acknowledges support from the FWF under Proj.~No. P26253-N27. LR acknowledges support from the European Commission through the Marie Sk\l{}odowska-Curie Fellowship No. 702298-DELTAS and the FWF through the Lise-Meitner Fellowship No. M 1650-N27. PvO acknowledges support from the FWF under Proj.~No. P27544-N28. 

\providecommand*{\mcitethebibliography}{\thebibliography}
\csname @ifundefined\endcsname{endmcitethebibliography}
{\let\endmcitethebibliography\endthebibliography}{}

\clearpage  
\begin{figure}[h] 
\centering
\includegraphics[width=\columnwidth, clip=true]{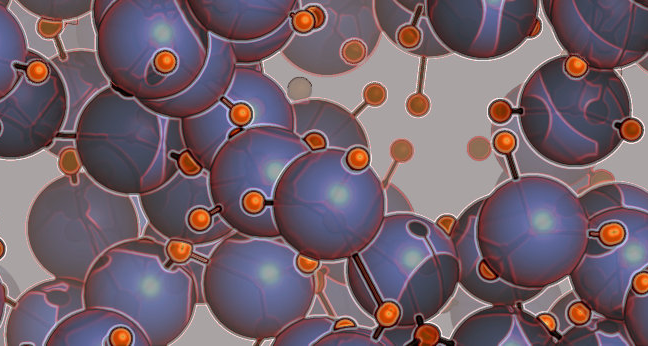}
\caption{{\bf Table of contents Figure}. Artistic representation of limited valance units consisting of a soft core (in blue) and a small number of flexible bonding patches (in orange).}
\label{fig:TabContents}
\end{figure}
 
\end{document}